\documentclass[runningheads]{llncs}
\usepackage[T1]{fontenc}
\usepackage{epsfig}
\usepackage{amssymb}
\usepackage{amsmath}
\usepackage{amsfonts}
\usepackage{float}
\usepackage{subcaption}
\usepackage{graphicx}
\usepackage{lscape}
\usepackage{rotating}
\usepackage{placeins}
\usepackage{multirow}
\usepackage{adjustbox}
\usepackage{tabularx}
\usepackage[colorlinks=true, linkcolor=blue, citecolor=blue, urlcolor=blue]{hyperref}
\begin{document}
\title{Inferring Prerequisite Knowledge Concepts in Educational Knowledge Graphs: A Multi-criteria Approach}
%
%
\author{Rawaa Alatrash\inst{1}\orcidID{0000-0003-2192-028X} \and
Mohamed Amine Chatti\inst{1}\orcidID{0000-0002-1311-7852} \and
Nasha Wibowo\inst{1}\and
Qurat Ul Ain\inst{1}\orcidID{0000-0003-2691-0267}
}
\authorrunning{R. Alatrash et al.}
%
\institute{Social Computing Group, Faculty of Computer Science, University of Duisburg-Essen, Duisburg, Germany}
%
\maketitle              
\begin{abstract}
Educational Knowledge Graphs (EduKGs) organize various learning entities and their relationships to support structured and adaptive learning. Prerequisite relationships (PRs) are critical in EduKGs for defining the logical order in which concepts should be learned. However, the current EduKG in the MOOC platform CourseMapper lacks explicit PR links, and manually annotating them is time-consuming and inconsistent. To address this, we propose an unsupervised method for automatically inferring concept PRs without relying on labeled data. We define ten criteria based on document-based, Wikipedia hyperlink-based, graph-based, and text-based features, and combine them using a voting algorithm to robustly capture PRs in educational content. Experiments on benchmark datasets show that our approach achieves higher precision than existing methods while maintaining scalability and adaptability, thus providing reliable support for sequence-aware learning in CourseMapper.
\keywords{Educational Knowledge Graphs \and MOOCs \and Prerequisite concept identification  \and Relation extraction \and Knowledge structure}
\end{abstract}
\section{Introduction}
Educational Knowledge Graphs (EduKGs) play a vital role in organizing and representing key entities and their relationships in digital learning environments, supporting knowledge discovery, concept navigation, and learner modeling \cite{ain2023automatic}. In our MOOC platform CourseMapper, EduKGs provide learners with a structured overview of the main knowledge concepts covered in learning materials to enhance understanding and engagement  \cite{ain2022learning,ALATRASH2024ConceptGCN,ain2023automatic}.
However, a major limitation of current EduKG structures in CourseMapper is the lack of prerequisite relationships (PRs), which define concept dependencies where understanding one concept is necessary before learning another. These relationships determine the logical sequence for mastering topics and support personalized sequence recommendations \cite{pan2017prerequisite}. Without explicit PRs, learners may struggle to identify which concepts should be mastered first, which can hinder self-directed learning and adaptive recommendation systems.
Prior research has explored end-to-end learning, feature-based, and hybrid methods for PR extraction \cite{gasparetti2018prerequisites,pan2017prerequisite,manrique2019exploring,bai2021bert}.  
End-to-end learning-based methods depend on domain-specific labeled data and static features, which limits their scalability. Feature-based methods remain widely used and rely on graph structures, text content, or Wikipedia data to capture concept dependencies \cite{manrique2019exploring,hu2022modeling}.
However, these methods have several limitations. They often overlook how semantic and structural information interact and they often rely only on concept order. Moreover, most feature-based approaches assess each criterion individually instead of combining their contributions for effective PR identification.
To overcome these limitations, we propose an unsupervised, multi-criteria approach that combines the contributions of document-based, Wikipedia hyperlink-based, graph-based, and text-based features using a voting algorithm, aiming to effectively extract PRs and enrich the structure of EduKGs in CourseMapper.
\section{Background and Related Work}
\subsection{EduKG Generation in CourseMapper}
\label{KG_generation}
Educational Knowledge Graphs (EduKGs) represent diverse entities and their relationships within a learning context, such as concepts, instructors, courses, and institutions \cite{ain2023automatic}. In our MOOC platform CourseMapper \cite{ain2022learning}, instructors can upload Learning Materials (LM), for which an EduKG is automatically generated. This includes entities such as LM, Slides (S), Main Concepts (MC), Related Concepts (RC), Categories (Cat), and Learners (L), linked by relationships such as CONTAINS, CONSISTS\_OF, RELATED\_TO, HAS\_CATEGORY, HAS\_READ, UNDERSTOOD (U), and DID\_NOT\_UNDERSTAND (DNU).
MCs are identified in each slide using a keyphrase extraction algorithm. These MCs undergo an expansion process via DBpedia Spotlight, which adds additional knowledge concepts as RCs and Cats. The RCs and Cats are then filtered to retain only the most relevant ones using cosine similarity in the EduKG\cite{ALATRASH2024ConceptGCN}. This structure provides learners with an organized view of key concepts in the LM.
The EduKG can be further enhanced by adding prerequisite relationships, which model the learning sequence among concepts and support adaptive learning pathways \cite{pan2017prerequisite}. Currently, CourseMapper’s EduKG does not include PRs, limiting its support for sequence-aware learning.
\subsection{Prerequisite Relationship Extraction}
PRs describe how concepts depend on each other in learning pathways. Methods for identifying PRs include end-to-end learning-based, feature-based, and hybrid approaches. End-to-end learning-based methods rely on deep learning to detect PRs from large datasets \cite{gasparetti2018prerequisites,xia2023course,thareja2023auto}. Feature-based methods use manually designed features \cite{pan2017prerequisite,manrique2019exploring,hu2022modeling} such as graph-based features \cite{manrique2019exploring}, which capture structural characteristics of graphs; text-based features, which use descriptions and content \cite{zhu2022predicting,hu2021active}; and Wikipedia-based features, which exploit hyperlinks, clickstreams, and categories \cite{wang2016using,bai2021bert,sayyadiharikandeh2019finding}. Hybrid methods combine both end-to-end learning-based and feature-based methods \cite{manrique2019exploring,bai2021bert}. 
Many existing end-to-end learning-based methods for PR inference depend on domain-specific labeled data and simplistic features, limiting scalability and adaptability. Manual annotation is time-consuming and inconsistent. Most feature-based methods consider criteria independently, often ignoring the interaction between semantic and structural information.
To overcome these issues, we propose an unsupervised multi-criteria approach combining ten distinct criteria document-based, Wikipedia hyperlink-based, graph-based, and text-based features.These criteria integrate semantic, structural, and contextual information through a \textbf{voting algorithm} that balances their contributions and mitigates noise from any single criterion.
Our method evaluates all concept pairs in the EduKG, which is crucial for unstructured content where structural criteria alone may be misleading. Therefore, incorporating additional criteria that leverage external resources and contextual information can enhance the accuracy of PR inference.
Our approach, tested on two public datasets, achieves a precision of $1.0$, demonstrating robustness, domain independence, and effective PR extraction.
\section{Methodology}
\label{methodology}
To define the likelihood of one concept being a prerequisite of another, we developed and adopted a set of criteria, derived from the literature review and adapted for our context. These criteria are categorized into document-based features, Wikipedia hyperlink-based features, graph-based features, hierarchy-based features, and text-based features.
Our multi-criteria approach begins by examining concept pairs against these criteria, each offering a unique perspective. Their outputs are then aggregated using a voting algorithm, followed by a thresholding step to infer whether a PR exists. These criteria are applied to all concept pairs in EduKG in CourseMapper and examined in both directions to ensure comprehensive identification. This study excludes concept synonyms and automatic redirections, as these are managed during the EduKG construction on the platform CourseMapper, ensuring only unique nodes without duplicates are extracted \cite{ALATRASH2024ConceptGCN}.
Before outlining the model structure, we adopt three core assumptions: \textbf{Transitivity}~\cite{bai2021bert,bai2025prerequisite}, where if $A \xrightarrow{\text{PR}} B$ and $B \xrightarrow{\text{PR}} C$, then $A \xrightarrow{\text{PR}} C$; \textbf{Asymmetry}~\cite{bai2021bert,bai2025prerequisite}, where if $B \rightarrow A$, then $A \nrightarrow B$; and \textbf{No Self-Connection}~\cite{bai2025prerequisite}, a concept cannot be a prerequisite of itself.
Table ~\ref{tab:Categories_of_Criteria} lists each criterion with its name, abbreviation, and description.
\begin{table}[ht]
\centering
\caption{Categories of Criteria with Full Names, Abbreviations, and Descriptions.}
\footnotesize
\begin{tabularx}{\textwidth}{|p{2.2cm}|p{3cm}|p{2.3cm}|X|}
    \hline
    \textbf{Category} & \textbf{Full name} & \textbf{Abbreviations} & \textbf{Description} \\
    \hline
    \shortstack{Document \\ based features} 
        & \shortstack{Temporal order} 
        & TemO 
        & \shortstack{$c_1$ precedes $c_2$  in the learning \\ material or channel} \\
    \hline
    \multirow{5}{*}{\shortstack{Wikipedia \\ Hyperlinks \\ based features}} 
        & \shortstack{Hyperlinks on \\ each other’s articles} 
        & HL-A  
        & \shortstack{$c_1$ hyperlink is mentioned \\ within $c_2$'s article} \\
    \cline{2-4}
        & \shortstack{Hyperlinks on \\ each other’s abstract} 
        & HL-Ab  
        & \shortstack{$c_1$ hyperlink appears in the\\ abstract of $c_2$} \\
    \cline{2-4}
        & \shortstack{Hyperlinks on each \\ other’s  related \\concepts’ abstracts} 
        & HL-RCA 
        & \shortstack{$c_1$ hyperlink is given on $c_2$'s \\related concept's abstract} \\
    \cline{2-4}
        & Reference distance 
        & RefD 
        & \shortstack{$c_1$ hyperlink is mentioned \\in the related concepts of $c_2$} \\
    \cline{2-4}
        & \shortstack{Inbound links and\\outbound links ratio} 
        & IOLR 
        & \shortstack{$c_1$ has higher ratio of the\\ number of in/outbound links \\ pointing to $c_2$} \\
    \hline
    \multirow{2}{*}{\shortstack{Graph \\ based features}} 
        & \shortstack{One concept is a \\category of another \\concept} 
        & CatCon
        & \shortstack{$c_1$ is a category of $c_2$ in \\DBPedia hierarchy} \\
    \cline{2-4}
        & \shortstack{One concept is the\\direct super-category \\of another concept} 
        & SuperCatCon
        & \shortstack{$c_1$ is a super category of  $c_2$ in\\ DBPedia hierarchy} \\
    \hline
    \shortstack{Hierarchy \\ based features} 
        & CourseMapper Hierarchy  
        & CMH 
        & \shortstack{first occurrence of $c_1$ in \\$channel_1$ precedes first\\occurrence of $c_2$ in $channel_2$ } \\
    \hline
    \shortstack{Text-\\based features} 
        & BERTopic and Shannon’s Entropy 
        & BERTropy 
        & \shortstack{$c_1$ has higher topic entropy \\than $c_2$} \\
    \hline
\end{tabularx}
\label{tab:Categories_of_Criteria}
\end{table}
\subsection{Prerequisite Relationships Extraction Features:}
\subsubsection{Document-based Features:}
 This category depends on the document structure to obtain the \textbf{temporal order} (TemO) of concepts within an article where the concept introduced first serves as a prerequisite for the subsequent \cite{xia2023course}.
For instance, in discussing \emph{Clustering}, the term \emph{machine learning} is likely introduced before \emph{Clustering} as understanding the former is essential for grasping the latter. \( TemO(c_1, c_2) \) is assigned the value $1$ if the position \( P(c_1) \) of concept \( c_1 \) appears before the position \( P(c_2) \) of concept \( c_2 \); otherwise, $0$.
\begin{equation} \label{equ:Temporal}
    TemO (c_1,c_2) = \begin{cases}    1,& \text{if } P(c_1)< P(c_2)\\
    0,              & \text{otherwise}
\end{cases}
\end{equation}
TemO applies only to MCs, as it relies on their order of appearance in the LM; related concepts are excluded because they are not explicitly mentioned.
\subsubsection{Wikipedia Hyperlinks-based Features:}
Wikipedia has been exploited to find potential PRs among concepts by analyzing the hyperlink structure of their corresponding Wikipedia articles \cite{xiao2022hybrid,bai2021bert}.

\textbf{Hyperlinks on each other's articles (HL-A)}:
In this criteria, we examine whether the hyperlinks for each concept appear in the corresponding Wikipedia articles of other concepts. For example, if the hyperlink for the concept \emph{Artificial Intelligence} ($c_1$) is found in the Wikipedia article for \emph{Machine Learning} ($c_2$), it suggests that \emph{Artificial Intelligence}  serves as a prerequisite for understanding \emph{Machine Learning}. This implies that $c_1$ is a prerequisite to $c_2$.
These relationships are formalized as follows:
\begin{equation}
    HL-A (c_1,c_2) = \begin{cases}    1,& \text{if } c_1 \in H_{c_2}\\
    0,              & \text{otherwise}
\end{cases}
\end{equation}
where $HL-A (c_1,c_2)$ determines whether the hyperlink for concept $c_1$ exists in the Wikipedia article of concept $c_2$, where in this case $1$ is assigned, otherwise $0$. $H_{c_2}$  represents the set of concepts that are hyperlinked within the Wikipedia article for concept $c_2$. 
Sometimes, hyperlinks in Wikipedia articles do not indicate a PR especially in long articles or hyperlinks at the end of the article.
For example, if \emph{Supervised Learning} appears at the end of the article \emph{Machine Learning} it could misleadingly suggest that \emph{Supervised Learning} is a prerequisite to \emph{Machine Learning}. Therefore, we need to consider criteria focused on the abstracts of articles for such special cases \cite{liang2018investigating}.

\textbf{Hyperlinks on each other's abstract (HL-Ab)}: This criterion is to find out if $c_1$ hyperlinks are included in $c_2$ abstract. The abstract typically provides a brief definition, and if hyperlink for concept $c_1$ is mentioned in the abstract of concept $c_2$, it suggests that $c_1$ is likely a prerequisite for $c_2$ \cite{liang2018investigating}. 
For instance, if \emph{Machine Learning} is referenced in the abstract of \emph{Supervised Learning} this implies that understanding \emph{Machine Learning} is necessary before learning \emph{Supervised Learning} \cite{liang2018investigating}.
The relationships are formulated as follows:
\begin{equation}
    HL-Ab (c_1,c_2) = \begin{cases}    1,& \text{if } {c_1} \in AH_{c_2}\\
    0,              & \text{otherwise}
\end{cases}
\end{equation}
$HL-Ab (c_1,c_2)$ indicates whether ${c_1}$ is a prerequisite to ${c_2}$, where $1$ is assigned in case hyperlink for concept ${c_1}$ belongs to $AH_{c_2}$ which is the set of concepts hyperlinked within the abstract of the Wikipedia article for concept $c_2$, otherwise $0$. 

 \textbf{Hyperlinks on each other's related concepts' abstracts (HL-RCA)}:
Inspired by the idea of analyzing the importance of related concepts to infer PRs \cite{liang2015measuring} and the role of hyperlinks in abstracts, we developed this criterion. This involves analyzing the abstracts of related concepts for each concept pair. For instance, consider the concepts \emph{artificial intelligence} and \emph{machine learning}. A related concept of \emph{machine learning}, such as \emph{artificial neural network}, is likely to reference a hyperlink for concept \emph{artificial intelligence} in the abstract of its Wikipedia article. In contrast, a concept related to \emph{artificial intelligence}, such as \emph{knowledge representation}, does not  mention the hyperlink of \emph{machine learning} concept within the abstract of its article. This pattern arises because related concepts often share common prerequisite concepts. In this example, \emph{artificial intelligence} serves as a prerequisite for both \emph{machine learning} and \emph{artificial neural network} as given in the following formula:
\begin{equation}
    HL-RCA (c_1,c_2) = \begin{cases}    1,& \text{if } c_1  \in A_{RC_{c_2}}\\
    0,              & \text{otherwise}
\end{cases}
\end{equation}
where $A_{RC_{c_2}}$ is the set of concepts that are hyperlinked within the abstract of the Wikipedia article of the related concepts of $c_2$. $HL-RCA(c_1,c_2)$ will be $1$ in case $c_1$ hyperlinks are present in $A_{RC_{c_2}}$, otherwise $0$.

 \textbf{Reference distance (RefD)}: It is defined as the reference distance between concept pair \cite{liang2015measuring}. This measures the proximity of Wikipedia articles based on their reference links and can be an excellent indicator for finding PRs. The idea here is that for each concept pair, we are comparing the asymmetry of $c_1$'s related concept mentioning $c_2$ and $c_2$'s related concepts mentioning $c_1$. For example, most of \emph{Data Mining} related concepts mention \emph{Data Science} in their articles. However, none of \emph{Data Science} related concepts mention \emph{Data Mining} in their articles. 
We adapted this equation to fit our system as follows:
\begin{equation} \label{eq:refd}
    RefD_{(c_1,c_2)} = 
    \frac{\sum^{k}_{i=1} r(R_{c_{i}},c_2) \, w(R_{c_{i}},c_1)}
         {\sum^{k}_{i=1} w(R_{c_{i}},c_1)}
    - \frac{\sum^{k}_{i=1} r(R_{c_{i}},c_1) \, w(R_{c_{i}},c_2)}
           {\sum^{k}_{i=1} w(R_{c_{i}},c_2)}
\end{equation}
where $C = \{c_1,...,c_k\}$ is the concept space, $R_{c_{i}}$ is a set of related concepts to $c_i$, $w(R_{c_{i}},c_1)$ is cosine similarity between $R_{c_{i}}$ and $c_1$, $r(R_{c_{i}},c_1)$ the relationship between related concept of $c_i$ and  $c_1$, where $R_{c_{i}}$ refers to $c_1$, meaning whether the related concept of $c_i$ includes $c_1$.
The outcome of RefD is then compared against a threshold to determine whether a PR is indicated as shown in the equation \ref{eq: refd2} below.
\begin{equation} \label{eq: refd2}
    RefD (c_1,c_2) = \begin{cases}    1,& \text{if }-1<RefD(c_1,c_2)<-\theta\\
    0,              & \text{otherwise}
\end{cases}
\end{equation}
$c_1$ is considered a prerequisite of $c_2$ if the RefD result falls between -1 and $-\theta$.
Conversely, $c_2$ is assumed to be a prerequisite of $c_1$ if the RefD result is between $\theta$ and 1. While between $-\theta$ and $\theta$ indicates no prerequisite can be inferred, where $\theta$ is empirically determined.
This is defined as the following:
\begin{equation}
    RefD (c_2,c_1) = \begin{cases}    1,& \text{if } \theta<RefD(c_1,c_2)<1\\
    0,              & \text{otherwise}
\end{cases}
\end{equation}
\textbf{Inbound links and outbound links ratio (IOLR)}:
This criterion evaluates a concept’s connectivity by comparing its inbound and outbound links \cite{liang2018investigating,bai2021bert}. Foundational concepts usually receive more inbound links, while advanced ones tend to have more outbound links. To reduce bias from generally popular concepts, we compute the inbound-to-outbound link ratio (IOL), where a higher IOL indicates that a concept is more foundational and likely to be a prerequisite, as given below:
\begin{equation}
    IOL(c_i) = \frac{\sum {inbound links}}{\sum {outbound links}}
\end{equation}
Then, we define the ratio to determine the direction of the relationship, as indicated by the equation $IOLR (c_1,c_2)$. 
\begin{equation}
    IOLR (c_1,c_2) = \begin{cases}    1,& \text{if } IOL(c_1) > IOL(c_2)\\
    0,              & \text{otherwise}
\end{cases}
\end{equation}
\subsubsection{Graph-based Features:}
This category adopts the hierarchical structure of DBpedia, where the categories in DBpedia represent the hierarchical relationships between concepts.

 \textbf{One concept is a category of another concept (CatCon)}:
Categories outline the essential topics of concepts, and understanding a concept often requires knowledge of its foundational elements defined by its category \cite{xiao2022hybrid,bai2021bert}. Based on the DBpedia hierarchy, if $c_2$ is classified under $c_1$, we can infer that $c_1$ is a category of $c_2$ and thus a prerequisite for $c_2$. 
To formulate this, we define $K_{c_2}$ as the category associated with $c_2$ and vice versa in case of $K_{c_1}$. This is given in the equation below:
\begin{equation}
    CatCon (c_1,c_2) = \begin{cases}    1,& \text{if } c_1 \in K_{c_2}\\
    0,              & \text{otherwise}
\end{cases}
\end{equation}
$CatCon(c_1,c_2)$ indicates that $c_1$ belongs to the category associated with $c_2$.  In other words, $c_2$ is categorized under $c_1$. This, strongly implies that $c_1$ serves as a prerequisite for $c_2$. 

\textbf{One concept is the direct super-category of another concept (SuperCatCon)}:
Another useful signal for identifying PRs is the direct super-category of a concept, defined as the category of its category in DBpedia \cite{xiao2022hybrid}. This follows a transitive logic: if 
$c_1$ \emph{(Machine Learning)} is a category of $c_2$ \emph{(Unsupervised Learning)}, and $c_2$ is a category of $c_3$ \emph{(K-Means)}. Thus, $c_1$ is the super-category of $c_3$, which means $c_1$ is inferred as a prerequisite of $c_3$.
Let $K^{up}_{c_1}$ denote the direct super-category of $c_1$, PR exists when these intersect with the categories of $c_2$.
\begin{equation}
    SuperCat(c_1,c_2) = |K^{up}_{c_1} \cap K_{c_2}|
\end{equation}
This is also repeated to calculate the existence of prerequisite in both directions $super(c_2,c_1)$. 
Then, we compare their asymmetry to avoid intertwining categories.
We want to prevent scenarios where some super-categories of $c_2$ are also categories of $c_1$ for $SuperCat(c_1,c_2)$ or for $SuperCat(c_2,c_1)$, which would imply mutual prerequisites, as follows: 
\begin{equation}
    SuperCatCon(c_1,c_2) = \begin{cases}    1,& \text{if } len(SuperCat(c_1,c_2)) 
 > len(SuperCat(c_2,c_1)) \\
    0,              & \text{otherwise}
\end{cases}
\end{equation}
When $SuperCatCon(c_1,c_2)$ is 1, indicates that more of $c_1$'s super-categories are also categories of $c_2$ suggesting that $c_1$ is a prerequisite of $c_2$. 
\subsubsection{ Hierarchy-based Features:}
This category captures the structural organization of educational content, reflecting concept relationships based on their positions within hierarchical formats. This approach can be applied to all educational platforms that follow a hierarchy structure. Inspired by Thareja et al., who modeled conceptual progression using the hierarchical structure of academic textbooks to detect PRs \cite{thareja2023auto}. We applied a similar strategy in our context by leveraging the course hierarchy in CourseMapper, to define the CourseMapper Hierarchy (\textbf{CMH}) criterion, which captures PRs among concepts within the course hierarchy. In CourseMapper, courses are structured hierarchically, encompassing multiple topics, each of which consists of several channels that, in turn, contain various LMs. This hierarchical organization can be leveraged to infer PRs, as the arrangement of topics, channels, and LMs follows a sequential order. Specifically, channels that need to be accessed earlier are positioned before those intended for later stages of learning. By exploiting this structured progression, we can systematically derive PRs between different concepts within one course. 
\begin{equation}
    ConTop (c_1,c_2) = \begin{cases}    1,& \text{if } Top(c_1) > Top(c_2)\\
    0,              & \text{otherwise}
\end{cases}
\end{equation}
In this context, \( Top(c_1) \) represents the topics or channels within one course in which \( c_1 \) appears within a learning material in CourseMapper. \( ConTop(c_1, c_2) \) is assigned a value of \( 1 \) if the channel in which \( c_1 \) is mentioned precedes the channel that mentions \( c_2 \). In this case, the likelihood of \( c_1 \) being a prerequisite for \( c_2 \) increases significantly. 
\subsubsection{Text-based Features:} This approach utilizes BERTopic combined with Shannon's entropy (\textbf{BERTropy}) to infer prerequisite connections ~\cite{bai2021bert}. In this context, a topic is defined as a theme discussed in the Wikipedia abstract of a given concept. For example, the abstract of the \emph{Machine Learning} article may include topics such as \emph{Artificial Intelligence} and \emph{Statistical Methods}. The underlying intuition of this approach is that more advanced concepts tend to be more specialized and therefore cover fewer topics, whereas general concepts span a broader set of topics. 
To operationalize this, BERTopic is employed to extract topics from the Wikipedia abstract of a concept $c$. The model assigns a probability to each topic, indicating the degree to which the topic is present in the abstract. Let $P(x)$ denote the probability assigned by BERTopic that topic $x$ appears in the abstract of concept $c$:
\begin{equation}
P(x) = \text{BERTopic}(x, c)
\end{equation}
Given these topic probabilities, Shannon entropy is used to measure the distribution of topics within the abstract:
\begin{equation}
    EntPro(c)=-\sum_x^X P(x)log \frac{1}{P(x)}
\end{equation}
where $X$ denotes the set of all identified topics and $P(x)$ represents the probability of topic $x$ for concept $c$. A higher entropy score indicates that the concept is more general, while a lower entropy score suggests a more advanced and focused concept.
This is applied to infer PRs between all concept pairs using entropy rule.
A PR is assumed to exist only if the difference exceeds this threshold $\theta$. In our case, the threshold is empirically selected and set 
$\theta= 1.33 $, meaning that between 0 and 1.33, we assume no PR exists.
\begin{equation}
    BERTropy(c_1,c_2) = \begin{cases}    1,& \text{if } EntPro(c_1)
 - EntPro(c_2) > \theta \\
    0,              & \text{otherwise}
\end{cases}
\end{equation}
\subsection{Voting Algorithm}
After applying the criteria mentioned previously to each concept pairs, a \textbf{voting algorithm} will take place determine both the presence and direction of the PR between concept pairs. First, we treat all criteria equally by assigning equal weight (1) to all of them. This avoids biasing inference toward any single feature and ensures no criterion dominates the final decision.
After that, we constructed two binary arrays (one hot encoded) ($A_1, A_2$) corresponds to ($c1 > c2$) and ($c2 < c1$), each of size 10, to examine how well each concept pair satisfies the set of predefined criteria. Each array element corresponds to a specific criterion and is assigned a value of 1 if the criterion is met or 0 if not. 
This means that only the criteria that are satisfied contribute to the final computation used to determine the PR, while the criteria that are not met have no influence on the result.
We then separately sum the positive values in each array to quantify the total contribution of satisfied criteria for each concept pair.
Following that, we compute the difference between the two sums by subtracting the total score of \( A_1 \) from that of \( A_2 \), as detailed in the following equation.
\begin{equation}
    S_{\text{init}} = \sum A_1 - \sum A_2
\end{equation}
After that, normalization takes place, which is a common practice in many applications and simplifies setting thresholds for PR inference. To this end, the final score ($S_{\text{init}}$) is rescaled from the original range of \( [-10, 10] \) to the standardized range of \( [-1,1] \). This facilitates meaningful aggregation and supporting the inference of PRs.
\begin{equation}
    S_{\text{new}} = 2 \cdot \left( \frac{S_{\text{init}} - S_{\text{min}}}{S_{\text{max}} - S_{\text{min}}} \right) - 1
\end{equation}
where $S_{new}$ is the normalized score difference between the two arrays, $A_1$ and $A_2$. $S_{min}$, in this case, would be the minimum, which is $-10$ and $S_{max}$  is the maximum, $10$. $S_{init}$ is the current score before normalization a concept pair has. After normalizing the results, we classify the results into three distinct ranges. To do this, we define a threshold on both sides, these are $\theta$ and $-\theta$. The first range is if $S_{new}$ or the difference between the arrays is between $-1$ and $-\theta$ which concludes that $c_2$ is a prerequisite concept to $c_1$. When the result is between $-\theta$ and $\theta$, that means there is no PR between the two concepts. The last range is when $S_{new}$ is between $\theta$ and $1$ then we conclude that $c_1$ is a prerequisite of $c_2$. The value of $\theta = 0.28$ is determined empirically through a series of experiments aimed at identifying the optimal parameter setting for our specific application as mentioned in Section \ref{Sec:Threshold_Optimization}. 
To illustrate this more, an example in Table \ref{table: example_array} is given, after examining all criteria for the concept pair \textit{Machine Learning} ($c_1$) and \textit{Supervised Learning} ($c_2$), we use two one-hot encoded arrays ($A_1, A_2$)  
to represent how well each concept pair satisfies a set of predefined criteria, in both directions.
\begin{table}
    \caption{Voting algorithm example}
    \label{table: example_array}
        \begin{tabular}{|l|c|c|c|c|c|c|c|c|c|c|}
        \hline
           Criteria  & TemO & HL-A & HL-Ab & HL-RCA & RefD & IOLR & CatCon & SuperCatCon & CMH & BERTropy \\
            \hline
            $A_1 $ {($c_1$->$c_2$)} &1 &1 &0 &0 &0 &0 &1 &1 &0 &1 \\
            \hline
           $A_2 $ {($c_2$->$c_1$)}  &0 &0 &1 &0 &1 &0 &0 &0 &0 &0\\
            \hline
        \end{tabular}
\end{table}
After that, we calculate the overall sums of the two arrays ($A_1 = 5.00$ and $A_2 =2.00$), then compute the difference between them and normalize it, as illustrated in Table~\ref{table: Normalised_values}. The final score of $0.3 > 0.28$ falls between $\theta$ and $1$, indicating that $c_1$ is prerequisite to $c_2$.
\begin{table}
\caption{Example of voting algorithm results}
\label{table: Normalised_values}
        \begin{tabular}{|l|c|c|c|c|}
            \hline
            Mechanism &Total $A_1$ ($c_1$->$c_2$) &Total $A_2$ ($c_2$->$c_1$) & $A_1 - A_2$ &Normalized difference \\
            \hline
            Voting algorithm &5.00 &2.00 &3.00 &0.3\\
            \hline
        \end{tabular}
\end{table}
\section{Experiments and Results}
\label{Experiments}
To demonstrate the effectiveness and advantages of our proposed method, we conducted thorough experiments and a comparative analysis using two public datasets, namely the AL-CPL \cite{wang2016using,liang2018active,liang2018investigating} and Biology datasets, which is a specialized dataset derived from textbook chapters in the Biology curriculum \cite{thareja2023auto}. 

\textbf{Threshold Optimization: }
\label{Sec:Threshold_Optimization}
The threshold is essentially a hyperparameter to control the sensitivity of the model in determining the PR. The thresholds of criteria RefD and BERTopic, as well as the proposed voting algorithm were determined empirically using the same methodology, based on the AL-CPL dataset. To this end, a range of threshold values have been used to run each approach separately and compared the results with the ground truth dataset to compute precision, accuracy, recall, and F1 score. 
The \textbf{RefD} method identifies PRs by analyzing how concepts reference each other within the Wikipedia link structure. Wang et al. \cite{wang2016using} originally proposed a threshold value of $\theta = 0.04$. We replicated this in the context of our platform, CourseMapper to generate the PRs. We found out that precision peaks at $\theta = 0.6$ with a value close to 0.9.
This aligned with our aim of reliably detecting PRs while minimizing false positives.
The higher threshold compared to previous work is attributed to the more domain-specific concept space used in our EduKG. While prior studies operated over the full Wikipedia concept set, our system limits the scope to concepts and related concepts extracted from AL-CPL, leading to a more focused and specialized semantic space.
Similarly, \textbf{BERTropy} infers PRs between concepts by analyzing the distribution of topics covered within their respective abstracts. We calculated the distribution of topics covered across all concepts. Then,  BERTropy is applied to the entire set of concepts to determine PRs among them.
We empirically found that precision peaks at $\theta = 1.4$, reaching approximately 0.9. This suggests that higher thresholds are more conservative, identifying only the strongest relationships, whereas lower thresholds connect more topics but may introduce noise. 
For the \textbf{Voting Algorithm}, we experimentally observed that precision reaches its maximum value of 1.0 at $\theta = 0.28$, indicating complete correctness among positive predictions without false positives at this threshold. 
In this study, we prioritize precision because minimizing false positives is more critical than capturing all possible PRs. Incorrect prerequisite suggestions could negatively affect learning paths and curriculum design.
\subsubsection{Experiment 1: Prerequisite Learning on the AL-CPL Dataset: }
To evaluate the effectiveness of our proposed model, we conducted experiments using three domains from the AL-CPL dataset: Data Mining, Physics, and Macroeconomics \cite{wang2016using,liang2018investigating,liang2018active}. It labels concepts with relationships: 1 for "prerequisite," -1 for "reverse prerequisite," and 0 for no relation. This dataset validates all criteria except CMH, which requires a content hierarchy that is lost when materials are uploaded into a single channel, treating all concepts at the same level. Therefore, hierarchical information was only applied in the Biology dataset, where such structure was available. The AL-CPL was uploaded in CourseMapper, and an EduKG was generated including main concepts, related concepts, and categories. This set of related concepts is utilized to infer the likelihood of a PR between concept pairs of the learning materials in CourseMapper.

\textbf{Evaluation of Individual Criteria:}
To evaluate the contribution of each criterion, we conducted experiments across three domains—Data Mining, Physics, and Macroeconomics. Results showed that BERTropy achieved (1), HL-Ab obtained (0.84), and RefD accomplished (0.84) notably enhanced precision in Data Mining, while HL-Ab got (1) and CatCon got (1) performed well in Physics, and CatCon excelled in Macroeconomics with value (0.86). In terms of recall, IOLR with value (0.83, 0.84, 0.91) consistently outperformed all other criteria across three domains (Data Mining, Physics, and Macroeconomics), effectively capturing a broader set of prerequisite relations. Conversely, SuperCatCon underperformed in both precision and recall, indicating the need for refinement.
To assess the balance between precision and recall, F1 scores were analyzed. IOLR achieved the highest F1 scores with values (0.67, 0.55, 0.59) in all domains (Data Mining, Physics, and Macroeconomics), confirming its effectiveness in maintaining this trade-off. Accuracy was also considered and offered useful trends, despite potential sensitivity to class imbalance. For example, HL-A showed high accuracy (0.65) in Data Mining, while CatCon achieved (0.70, 0.63) strong accuracy in Physics and Macroeconomics but had lower F1 scores, suggesting it may favor majority classes while poor at finding the minority class. The variation in criterion performance across domains can be attributed to differences in concept structure and semantic clarity. For instance, Data Mining and Physics datasets often contain well-defined, hierarchical, and technically specific concepts, which align well with criteria that leverage structural or semantic features. In contrast, the Macroeconomics domain tends to be more abstract and less formally structured, making it harder for some methods to identify clear PRs.
In line with prior work, we place greater emphasis on precision, as the reliability of inferred PRs is critical for downstream tasks and system utility. 

\textbf{Comparative Results and Analysis:}
 We compared our proposed model against six baseline methods—RefD \cite{liang2015measuring}, Active Learning \cite{liang2018active}, AdaBoost \cite{zhou2020ensemble}, WikiCPRL \cite{xiao2023wikicprl}, Neural Network \cite{moggio2020unige}, and a Linguistically-Driven Strategy \cite{miaschi2019linguistically}—to evaluate its effectiveness in capturing PRs without relying on labeled data. As shown in Table ~\ref{table:results_no_biology}, we assessed our voting algorithm across three domains: Data Mining, Physics, and Macroeconomics.
Compared to existing baselines, our proposed Voting Algorithm consistently achieves higher precision but lower recall and F1 scores, reflecting its focus on reliably extracting only the strongest edges. For example, in Data Mining, our method reaches perfect precision (1.0) while RefD, Active Learning, and AdaBoost achieve lower precision (0.51,0.80, 0.85) but higher recall (0.76, 0.73, 0.85) and F1 scores (0.61, 0.67, 0.85). Similarly in Physics, Bagging-based AdaBoost achieves higher precision (0.87) than our method (0.76), while other baselines such as RefD and WikiCPRL show lower precision (0.49 and 0.43, respectively) compared to our method, but higher recall (0.49, 0.90).
The same trend is observed for the Neural Network and the Linguistically-Driven Strategy, which yield higher F1 scores (0.65 and 0.81 in Data Mining, respectively, vs. 0.42 for our method) due to their stronger recall. This pattern shows that while baselines recover more links overall, our method prioritizes the precision and reliability of inferred PRs—critical for building robust learning paths—since, in educational settings, misleading a learner with an incorrect prerequisite (false positive) is more harmful than omitting some prerequisites.

However, baseline models perform better in recall, which we attribute to our precision-focused threshold selection in criteria such as RefD and BERTropy. These thresholds were tuned to prioritize reliability over quantity, resulting in fewer but more accurate inferred relationships. This can be attributed to differences between domains in concept structure and semantic clarity as mentioned earlier. 
Despite the lower recall, the proposed voting algorithm which integrates various feature-based methods achieves competitive performance without requiring any labeled training data. This makes it particularly well-suited for real-world settings where labeled prerequisite data is limited or difficult to obtain.
Although supervised models often outperform unsupervised ones due to access to labeled data, manual labeling is costly, time-consuming, and prone to inconsistency. Most unsupervised methods evaluate criteria independently rather than integrating them for more effective PR identification. Our unsupervised multi-criteria approach addresses this gap with a scalable, flexible voting algorithm for inferring PRs, particularly in diverse educational domains lacking annotations.
\begin{table}
    \centering
    \caption{Performance Comparison Across AL-CPL Datasets}
    \label{table:results_no_biology}
    \resizebox{\textwidth}{!}{
        \begin{tabular}{|l|c|c|c|c|c|c|c|c|c|c|c|c|} 
            \hline 
            & \multicolumn{4}{|c|}{Data Mining} & \multicolumn{4}{|c|}{Physics} & \multicolumn{4}{|c|}{Macroeconomics} \\ 
            \hline 
            Criteria & P & R & F1 & Acc. & P & R & F1 & Acc. & P & R & F1 & Acc. \\ 
            \hline 
            TemO & 0.47 & 0.31 & 0.37 & 0.40 & 0.36 & 0.42 & 0.39 & 0.47 & 0.46 & 0.71 & 0.56 & 0.53 \\ 
            \hline 
            HL-A & 0.8 & 0.53 & 0.64 & \textbf{0.65} & 0.58 & 0.34 & 0.43 & 0.64 & 0.58 & 0.34 & 0.42 & 0.62 \\ 
            \hline 
            HL-Ab & 0.84 & 0.22 & 0.35 & 0.53 & \textbf{1} & 0.13 & 0.23 & 0.65 & 0.62 & 0.08 & 0.14 & 0.60 \\ 
            \hline 
            HL-RCA & 0.53 & 0.53 & 0.53 & 0.45 & 0.42 & 0.67 & 0.52 & 0.49 & 0.39 & 0.43 & 0.41 & 0.48 \\ 
            \hline 
            RefD & 0.84 & 0.22 & 0.35 & 0.53 & 0.72 & 0.21 & 0.33 & 0.65 & 0.69 & 0.15 & 0.24 & 0.62 \\ 
            \hline 
            IOLR & 0.56 & \textbf{0.83} & \textbf{0.67} & 0.53 & 0.41 & \textbf{0.84} & \textbf{0.55} & 0.45 & 0.44 & \textbf{0.91} & \textbf{0.59} & 0.47 \\ 
            \hline  
            CatCon & 0 & 0 & 0 & 0.43 & \textbf{1} & 0.25 & 0.40 & \textbf{0.70} & \textbf{0.86} & 0.12 & 0.21 & \textbf{0.63} \\ 
            \hline 
            SuperCatCon & 0 & 0 & 0 & 0.43 & 0 & 0 & 0 & 0.60 & 0 & 0 & 0 & 0.59 \\ 
            \hline 
            BERTropy & \textbf{1} & 0.09 & 0.16 & 0.48 & 0.26 & 0.09 & 0.13 & 0.54 & 0 & 0 & 0 & 0.59 \\ 
            \hline
            \hline
            \multicolumn{13}{|c|}{\textbf{Baselines}} \\ 
            \hline 
            RefD \cite{liang2015measuring} & 0.51 & 0.76 & 0.61 & - & 0.49 & 0.49 & 0.49 & - & - & - & - & - \\ 
            \hline 
            Active Learning \cite{liang2018investigating} & 0.80 & 0.73 & 0.76 & - & 0.8 & 0.6 & 0.7 & - & - & - & - & - \\ 
            \hline 
            Bagging-based AdaBoost \cite{zhou2020ensemble} & 0.85 & 0.85 & \textbf{0.85} & - & \textbf{0.87} & \textbf{0.87} & \textbf{0.86} & - & - & - & - & - \\ 
            \hline 
            WikiCPRL \cite{xiao2023wikicprl} & 0.59 & \textbf{0.90} & 0.68 & - & 0.43 & 0.83 & 0.57 & - & - & - & - & - \\ 
            \hline 
            Neural Network \cite{moggio2020unige} & 0.62 & 0.68 & 0.65 & - & 0.53 & 0.62 & 0.57 & - & - & - & - & - \\ 
            \hline 
            Linguistically-Driven Strategy \cite{miaschi2019linguistically} & 0.81 & 0.81 & 0.81 & - & 0.77 & 0.71 & 0.74 & - & - & - & - & - \\ 
            \hline 
            \textit{Proposed -  Voting Algorithm} & \textbf{1} & 0.27 & 0.42 & 0.58 & 0.76 & 0.13 & 0.22 & 0.64 & 0.65 & 0.16 & 0.25 & 0.61 \\ 
            \hline
        \end{tabular}
    }  
\end{table}
\subsubsection{Experiment 2: Prerequisite Learning on the Biology Dataset: }
\label{Experiement_2}
To effectively evaluate the proposed model’s ability to detect PRs and assess \textbf{TemO} criterion more reliably, as well as to measure the contribution of \textbf{CMH}, we utilized a specialized dataset derived from textbook chapters in the NCERT\footnote{https://ncert.nic.in/textbook.php} Biology curriculum \cite{thareja2023auto}. 
For our experiments, we selected two conceptually related chapters from NCERT science textbooks: Chapter 7 (Transportation in Animals and Plants, Class 7) and Chapter 6 (Life Processes, Class 10). These were chosen for their natural hierarchical relationship, with foundational content in Class 7 supporting advanced concepts in Class 10.
Both chapters were uploaded into CourseMapper under the same course and same topic, but placed in two different sequential channels to preserve the learning hierarchy defined by the platform. We generated the EduKG for each chapter individually by extracting the main concepts from each and then expanding them using DBpedia to identify related concepts. This structure enabled us to evaluate all ten predefined criteria for detecting PRs. It is worth mentioning, when a new learning material is added, the system computes the EduKG for this new material, without needing to reprocess the entire EduKG. 
In the absence of ground truth in the original dataset, 186 concept pairs were manually annotated by domain experts (PhDs and medical doctors) as either prerequisite or not. 

\textbf{Evaluation of Individual Criteria and Comparative Results:}
Biology dataset is used to evaluate each individual criterion using four standard metrics: precision, recall, F1-score, and accuracy (Table ~\ref{table:results_biology}).
HL-Ab and HL-A achieved the highest precision, effectively identifying the most reliable PRs while minimizing false positives. In contrast, IOLR and CMH attained the highest recall, capturing a broader set of dependencies at the expense of tolerating some false positives. Notably, IOLR and CMH also achieved the highest F1-scores, indicating a strong balance between precision and recall.
In terms of accuracy, BERTropy performed best (0.75), followed closely by IOLR and CMH, reinforcing their robustness in distinguishing between prerequisite and non-prerequisite pairs. The success of CMH, in particular, highlights the value of structured educational content and hierarchical information in inferring PRs.
In this dataset, the placement of materials across channels is well-defined, enabling CMH to leverage this ordering effectively. By contrast, CatCon and SuperCatCon depend on the category hierarchy in DBpedia, and TemO relies on the sequential appearance of concepts in text, both of which are less consistent in the current dataset. Their effectiveness may improve by incorporating additional chapters.
In comparison, our proposed voting algorithm achieved competitive results, although it did not surpass the best-performing individual criteria such as CMH and IOLR. With moderate precision and recall, it provided a balanced yet limited performance compared to these criteria when applied independently.
The effectiveness of the voting algorithm is largely supported by strong individual contributors like CMH and IOLR. CMH achieved high F1-scores due to its reliance on hierarchical signals, while IOLR benefited from the ratio of in-outbound links—both features being particularly well captured in the Biology dataset. These findings highlight the importance of integrating hierarchical information. The defined criteria are deduced from Wikipedia articles, DBpedia hierarchy, textual information, content order, and platform hierarchy to infer PRs, making it applicable across diverse educational platforms.
\begin{table}
    \centering
    \caption{Performance Comparison Across Biology dataset}
    \label{table:results_biology}
        \begin{tabular}{|l|c|c|c|c|} 
            \hline 
            & \multicolumn{4}{|c|}{Biology} \\ 
            \hline 
            Criteria & P & R & F1 & Acc. \\ 
            \hline 
            TemO & 0.0 & 0.0 & 0.0 & 0.42 \\ 
            \hline 
            HL-A & 0.91 & 0.32 & 0.48 & 0.58 \\ 
            \hline 
            HL-Ab & \textbf{1.0} & 0.06 & 0.12 & 0.45 \\ 
            \hline 
            HL-RCA & 0.87 & 0.23 & 0.35 & 0.51 \\ 
            \hline 
            RefD & 0.75 & 0.10 & 0.17 & 0.45 \\ 
            \hline 
            IOLR & 0.58 & \textbf{1.0} & \textbf{0.74} & 0.58 \\ 
            \hline 
            CMH & 0.58 & \textbf{1.0} & \textbf{0.74} & 0.58 \\ 
            \hline 
            CatCon & 0.0 & 0.0 & 0.0 & 0.42 \\ 
            \hline 
            SuperCatCon & 0.0 & 0.0 & 0.0 & 0.42 \\ 
            \hline 
            BERTropy & 0.55 & 0.65 & 0.70 & \textbf{0.75} \\ 
            \hline 
            \textit{Proposed - Voting Algorithm} & 0.48 & 0.35 & 0.41 & 0.40 \\ 
            \hline
        \end{tabular}
\end{table}
\section{Conclusion and Future Work}
Educational Knowledge Graphs (EduKGs) capture diverse entities and connections in educational systems, providing a structured view of key concepts from learning materials. Prerequisite relationships (PRs) are essential for defining learning sequences but are often missing in EduKGs. However, automatic methods for PR identification depend on domain-specific labeled data and static features, limiting scalability. Existing feature-based methods often ignore how semantic and structural information interact, rely often on concept order, and assess criteria individually. To address these gaps, we proposed an unsupervised method that combines ten criteria using a voting algorithm to infer PRs without using labeled data. Experiments on two benchmark datasets show that our proposed approach achieves higher precision than baselines and reliably extracts PRs across domains. In future work, we aim to extend our PR identification framework with more criteria. We will also experiment with hybrid and LLM-based approaches for more effective PR identification. Additionally, adaptive thresholding methods, learned per domain and combined with a weighting approach, could be adopted to further enhance the robustness and accuracy of PR inference.
\begin{credits}
\subsubsection{\discintname}
The authors have no competing interests to declare that are
relevant to the content of this article. 
\end{credits}
%
%
%
\bibliographystyle{splncs04}
\bibliography{sigproc}

\begin{thebibliography}{10}
\providecommand{\url}[1]{\texttt{#1}}
\providecommand{\urlprefix}{URL }
\providecommand{\doi}[1]{https://doi.org/#1}

\bibitem{ain2023automatic}
Ain, Q.U., Chatti, M.A., Bakar, K.G.C., Joarder, S., Alatrash, R.: Automatic construction of educational knowledge graphs: a word embedding-based approach. Information  \textbf{14}(10), ~526 (2023)

\bibitem{ain2022learning}
Ain, Q.U., Chatti, M.A., Joarder, S., Nassif, I., Wobiwo~Teda, B.S., Guesmi, M., Alatrash, R.: Learning channels to support interaction and collaboration in coursemapper. In: Proceedings of the 14th International Conference on Education Technology and Computers. pp. 252--260 (2022)

\bibitem{ALATRASH2024ConceptGCN}
Alatrash, R., Chatti, M.A., Ain, Q.U., Fang, Y., Joarder, S., Siepmann, C.: Conceptgcn: Knowledge concept recommendation in moocs based on knowledge graph convolutional networks and sbert. Computers and Education: Artificial Intelligence  \textbf{6},  100193 (2024). \doi{https://doi.org/10.1016/j.caeai.2023.100193}, \url{https://www.sciencedirect.com/science/article/pii/S2666920X23000723}

\bibitem{bai2025prerequisite}
Bai, Y., Liu, Z., Guo, T., Hou, M., Xiao, K.: Prerequisite relation learning: A survey and outlook. ACM Computing Surveys  (2025)

\bibitem{bai2021bert}
Bai, Y., Zhang, Y., Xiao, K., Lou, Y., Sun, K.: A bert-based approach for extracting prerequisite relations among wikipedia concepts. Mathematical Problems in Engineering  \textbf{2021}(1),  3510402 (2021)

\bibitem{gasparetti2018prerequisites}
Gasparetti, F., De~Medio, C., Limongelli, C., Sciarrone, F., Temperini, M.: Prerequisites between learning objects: Automatic extraction based on a machine learning approach. Telematics and Informatics  \textbf{35}(3),  595--610 (2018)

\bibitem{hu2022modeling}
Hu, H., Pan, L., Ran, Y., Kan, M.Y.: Modeling and leveraging prerequisite context in recommendation. arXiv preprint arXiv:2209.11471  (2022)

\bibitem{hu2021active}
Hu, X., He, Y., Sun, G.: Active learning for concept prerequisite learning in wikipedia. In: Proceedings of the 2021 13th International Conference on Machine Learning and Computing. pp. 582--587 (2021)

\bibitem{liang2015measuring}
Liang, C., Wu, Z., Huang, W., Giles, C.L.: Measuring prerequisite relations among concepts. In: Proceedings of the 2015 conference on empirical methods in natural language processing. pp. 1668--1674 (2015)

\bibitem{liang2018investigating}
Liang, C., Ye, J., Wang, S., Pursel, B., Giles, C.L.: Investigating active learning for concept prerequisite learning. In: Proceedings of the AAAI Conference on Artificial Intelligence. vol.~32 (2018)

\bibitem{liang2018active}
Liang, C., Ye, J., Zhao, H., Pursel, B., Giles, C.L.: Active learning of strict partial orders: A case study on concept prerequisite relations. arXiv preprint arXiv:1801.06481  (2018)

\bibitem{manrique2019exploring}
Manrique, R., Pereira, B., Mari{\~n}o, O.: Exploring knowledge graphs for the identification of concept prerequisites. Smart Learning Environments  \textbf{6}(1), ~21 (2019)

\bibitem{miaschi2019linguistically}
Miaschi, A., Alzetta, C., Cardillo, F.A., Dell’Orletta, F.: Linguistically-driven strategy for concept prerequisites learning on italian. In: Proceedings of the Fourteenth Workshop on Innovative Use of NLP for Building Educational Applications. pp. 285--295 (2019)

\bibitem{moggio2020unige}
Moggio, A., Parizzi, A.: Unige se@ prelearn: Utility for automatic prerequisite learning from italian wikipedia. EVALITA Evaluation of NLP and Speech Tools for Italian-December 17th, 2020 p.~376 (2020)

\bibitem{pan2017prerequisite}
Pan, L., Li, C., Li, J., Tang, J.: Prerequisite relation learning for concepts in moocs. In: Proceedings of the 55th Annual Meeting of the Association for Computational Linguistics (Volume 1: Long Papers). pp. 1447--1456 (2017)

\bibitem{sayyadiharikandeh2019finding}
Sayyadiharikandeh, M., Gordon, J., Ambite, J.L., Lerman, K.: Finding prerequisite relations using the wikipedia clickstream. In: Companion Proceedings of The 2019 World Wide Web Conference. pp. 1240--1247 (2019)

\bibitem{thareja2023auto}
Thareja, R., Garg, R., Baghel, S., Dwivedi, D., Mohania, M., Kulshrestha, R.: Auto-req: Automatic detection of pre-requisite dependencies between academic videos. In: Proceedings of the 18th Workshop on Innovative Use of NLP for Building Educational Applications (BEA 2023). pp. 539--549 (2023)

\bibitem{wang2016using}
Wang, S., Ororbia, A., Wu, Z., Williams, K., Liang, C., Pursel, B., Giles, C.L.: Using prerequisites to extract concept maps from textbooks. In: Proceedings of the 25th acm international on conference on information and knowledge management. pp. 317--326 (2016)

\bibitem{xia2023course}
Xia, J., Li, M., Tang, Y., Yang, S.: Course map learning with graph convolutional network based on aucm. World Wide Web  \textbf{26}(5),  3483--3502 (2023)

\bibitem{xiao2022hybrid}
Xiao, K., Fu, Y., Zhang, J., Tianji, W.: A hybrid approach for discovering concept prerequisite relations in wikipedia. In: 2022 9th International Conference on Behavioural and Social Computing (BESC). pp.~1--5. IEEE (2022)

\bibitem{xiao2023wikicprl}
Xiao, K., Li, K., Zhang, Y., Chen, X., Lou, Y.: Wikicprl: A weakly supervised approach for wikipedia concept prerequisite relation learning. In: Asia-Pacific Web (APWeb) and Web-Age Information Management (WAIM) Joint International Conference on Web and Big Data. pp. 177--192. Springer (2023)

\bibitem{zhou2020ensemble}
Zhou, Y., Xiao, K., Zhang, Y.: An ensemble learning approach for extracting concept prerequisite relations from wikipedia. In: 2020 16th International Conference on Mobility, Sensing and Networking (MSN). pp. 642--647. IEEE (2020)

\bibitem{zhu2022predicting}
Zhu, Y., Zamani, H.: Predicting prerequisite relations for unseen concepts. In: Proceedings of the 2022 Conference on Empirical Methods in Natural Language Processing (EMNLP 2022) (2022)

\end{thebibliography}
%




\end{document}